\documentclass[pra,showpacs,twocolumn,superscriptaddress]{revtex4-1}
\usepackage{mathrsfs}
\usepackage{amsfonts}
\usepackage{amsmath}
\usepackage{txfonts}
\usepackage{amssymb}
\usepackage{graphicx}
\usepackage{bm}
\usepackage{color}
\usepackage[normalem]{ulem}

\newcommand{\ket}[1]{|#1\rangle}
\newcommand{\bra}[1]{\langle #1|}
\newcommand{\Tr}{\text{Tr}}
\begin{document}
\title{{\it i}SWAP-type geometric gates induced by paths on Schmidt sphere}
\author{Max Johansson Saarij\"arvi}
\affiliation{Department of Physics and Astronomy, Uppsala University,
Box 516, Se-751 20 Uppsala, Sweden}
\author{Erik Sj\"oqvist}
\email{erik.sjoqvist@physics.uu.se}
\affiliation{Department of Physics and Astronomy, Uppsala University,
Box 516, Se-751 20 Uppsala, Sweden}
\date{\today}
\begin{abstract}
We propose {\it i}SWAP-type quantum gates based on geometric phases purely associated 
with paths on the Schmidt sphere [Phys. Rev. A {\bf 62}, 022109 (2000)]. These {\it geometric 
Schmidt gates} can entangle qubit pairs to an arbitrary degree; in particular, they can 
create  maximally entangled states from product states by an appropriate choice of  
base point on the Schmidt sphere. We identify Hamiltonians that generate pure paths 
on the Schmidt sphere by reverse engineering and demonstrate explicitly that the resulting 
Hamiltonians can be implemented in systems of transmon qubits. The geometric Schmidt 
gates are characterized by vanishing dynamical phases and are  complementary to 
geometric single-qubit gates that take place on the Bloch sphere.  
\end{abstract}
\maketitle
\date{\today}
\section{Introduction}
Geometric quantum computation \cite{ekert00,zhang23} is the idea to use Abelian 
geometric phases \cite{berry84,aharonov87} to build robust quantum gates. This has 
been implemented on different experimental platforms, such as nuclear magnetic 
resonance \cite{jones00,wang07}, trapped ions \cite{kim08}, electron spin resonance 
\cite{wu13}, NV centers in diamond \cite{kleissler18}, semiconducting spin and 
charge qubits \cite{zhang20,zhang21}, superconducting qubits \cite{xu20,setiawan23}, 
and Rydberg atoms \cite{xu22,su23}. A universal set of geometric gates requires arbitrary 
single-qubit gates solely dependent upon paths on the Bloch sphere, supplemented 
by a geometric gate that can entangle pairs of qubits \cite{zhu03}. 

Realizations of two-qubit gates are particularly challenging as they are limited by the 
naturally occurring type of qubit-qubit interaction in the chosen system \cite{schuch03}. 
For instance, while entangling gates such as CNOT and controlled phase flip can be 
implemented using a single application of Ising  interaction terms, {\it i}SWAP-type entangling 
gates such as $i$SWAP and $\sqrt{{\rm SWAP}}$ are similarly implementable in the 
presence of various forms of spin exchange interactions, such as XY or Heisenberg 
\cite{schuch03}. As interaction terms of these types are common in several qubit systems 
\cite{tanamoto09,rasmussen20}, it becomes pertinent to develop schemes for 
geometric {\it i}SWAP-type gates. Here, we propose such a general approach to 
geometric two-qubit gates. 

The idea of our proposal is based on the Schmidt decomposition \cite{nielsen00}, 
i.e., that any pure bipartite state can be written on Schmidt form, being a superposition 
of Schmidt vectors. These vectors are products of mutually orthogonal states for 
each subsystem.  When the two subsystems are qubits, the Schmidt decomposition 
can be represented on a two-dimensional `Schmidt sphere' \cite{sjoqvist00}. This 
sphere is parametrized by a polar angle that determines the degree of qubit-qubit 
entanglement and an azimuthal angle that takes care of the relative phase of the 
Schmidt vectors. The relevance of the Schmidt sphere has been demonstrated 
experimentally using polarization-entangled photon pairs \cite{loredo14}. 

Here, we introduce {\it geometric Schmidt gates} that are two-qubit gates controlled 
by the solid angle enclosed on the Schmidt sphere. This is achieved by designing  a 
complete set of orthonormal two-qubit states that acquire no dynamical phase and 
whose Schmidt vectors are kept constant throughout the implementation of the gate. 
These gates are the two-qubit analog of the single-qubit geometric gates associated 
with paths on the Bloch sphere. The geometric Schmidt gates are in a sense optimal 
as they require the same amount of parameter control as their single-qubit counterparts. 
We examine their ability to entangle the qubit pair by analyzing Makhlin's local invariants 
\cite{makhlin02}. We  identify the Hamiltonians that generate the gates by means of 
reverse engineering \cite{berry09} and demonstrate that they can be realized in systems 
with controllable spin exchange terms. 

\section{Schmidt gates}
Consider two qubits with local Hilbert spaces $\mathscr{H}_a$ and $\mathscr{H}_b$. 
Any pure state belonging to $\mathscr{H}_a \otimes \mathscr{H}_b$ of the two qubits 
can, up to an unimportant overall phase, be written on Schmidt form \cite{nielsen00}:
\begin{eqnarray}
\ket{\Gamma_+ ({\bf r})} = f \ket{{\bf n}} \otimes \ket{{\bf m}} + 
g \ket{-{\bf n}} \otimes \ket{-{\bf m}} , 
\end{eqnarray}
where $\ket{\pm {\bf n}}$ and $\ket{\pm {\bf m}}$ are orthonormal vector pairs 
belonging to $\mathscr{H}_a$ and $\mathscr{H}_b$, respectively. The 
amplitudes $f = e^{-i\beta/2} \cos \frac{\alpha}{2}$ and $g = e^{i\beta/2} \sin \frac{\alpha}{2}$ 
define the angles $\alpha$ and $\beta$ that parameterize a point  
\begin{eqnarray}
{\bf r} = (\sin \alpha \cos \beta,\sin \alpha \sin \beta , \cos \alpha)  
\end{eqnarray}
on the Schmidt sphere \cite{sjoqvist00}. Note that only the polar angle $\alpha$ is related 
to the amount of entanglement, as the azimuthal angle $\beta$ can be controlled by locally 
manipulating one of the qubits. The Bloch vectors of the reduced density operators of 
the two qubits point along ${\bf n}$ and ${\bf m}$ in the case where $|f| \neq |g|$, while 
these vectors distinguish different maximally entangled states \cite{milman03,milman06} 
when $|f| = |g|$ \cite{remark1}. Thus, any two-qubit state is fully specified by the triplet 
$({\bf r},{\bf n},{\bf m})$ and the evolution of the system can be viewed as 
paths on the local Bloch spheres and the Schmidt sphere \cite{loredo14}. 

In order to implement geometric Schmidt gates, we supplement $\ket{\Gamma_{+} ({\bf r})}$ 
with the states 
\begin{eqnarray}
\ket{\Gamma_{-} ({\bf r})} & = &  
-g^{\ast} \ket{{\bf n}} \otimes \ket{{\bf m}} + 
f^{\ast} \ket{-{\bf n}} \otimes \ket{-{\bf m}} , 
\nonumber \\ 
\ket{\Lambda_{+}} & = &  
\ket{{\bf n}} \otimes \ket{-{\bf m}} , \ \ \ \ 
\ket{\Lambda_{-}} =  
\ket{-{\bf n}} \otimes \ket{{\bf m}}
\label{eq:schmidtvectors}
\end{eqnarray}
and require ${\bf n}\equiv {\bf n}_0$ and  ${\bf m} \equiv {\bf m}_0$ to be fixed throughout 
the implementation of the gate. Thus, the Schmidt gate scenario is the special case where 
the evolution path is nontrivial only on the Schmidt sphere. Provided the dynamical phases 
all vanish or can be factored out as an overall global phase, a loop $\mathscr{C}:[0,\tau] \ni t 
\mapsto {\bf r}_t$, ${\bf r}_{\tau} = {\bf r}_0 \equiv ( \sin \alpha_0 \cos \beta_0 , \sin \alpha_0 
\sin \beta_0 , \cos \alpha_0 )$, that encloses a solid angle $\Omega$ on the Schmidt sphere 
induces the geometric two-qubit gate 
\begin{eqnarray}
 & & {\rm U}_{\alpha_0,\beta_0} (\mathscr{C}) = 
\ket{{\bf n}_0} \bra{{\bf n}_0} \otimes \ket{-{\bf m}_0} \bra{-{\bf m}_0} 
\nonumber \\ 
 & & + 
\ket{-{\bf n}_0} \bra{-{\bf n}_0} \otimes \ket{{\bf m}_0} \bra{{\bf m}_0} + 
e^{-i\Omega /2} \ket{\Gamma_{+} ({\bf r}_0)}\bra{\Gamma_{+} ({\bf r}_0)} 
 \nonumber \\
 & & + e^{i\Omega /2} \ket{\Gamma_{-} ({\bf r}_0)}\bra{\Gamma_{-} ({\bf r}_0)} .
\end{eqnarray}
This is the desired geometric Schmidt gate whose action is controlled by $\Omega$ 
enclosed by the loop $\mathscr{C}$ on the Schmidt sphere. 

\subsection{Entangling capability}
The ability of ${\rm U}_{\alpha_0,\beta_0} (\mathscr{C})$ to entangle the qubit pair relies 
on the base point ${\bf r}_0$ on the Schmidt sphere. To see this, let us consider two 
extreme cases: (i) ${\bf r}_0 = (0,0,1)$ and (ii) ${\bf r}_0 = (0,1,0)$. We shall see that 
while (i) cannot entangle, (ii) contains, up to a rotation around the $z$ axis, the only 
special perfect entangler \cite{rezakhani04}, i.e., the only geometric Schmidt gate that 
can create maximally entangled states from an orthonormal basis of product states. 
In the following, we put ${\bf n}_0 = -{\bf m}_0 = (0,0,1)$ such that $\ket{{\bf n}_0} = \ket{0}$ 
and $\ket{{\bf m}_0} = \ket{1}$, as well as use short-hand notation $\ket{xy} \equiv 
\ket{x}\otimes \ket{y}$, $x,y=0,1$, and $AB\equiv A\otimes B$ for operators $A$ 
and $B$ acting on $\mathscr{H}_a$ and $\mathscr{H}_b$, respectively. 

In case (i), we find the product gate 
\begin{eqnarray}
{\rm U}^{(i)} (\mathscr{C}) & \equiv & {\rm U}_{0,0} (\mathscr{C}) = \ket{00} \bra{00} + 
\ket{11} \bra{11}  
\nonumber \\ 
 & & + e^{-i\Omega /2} \ket{01}\bra{01} + 
 e^{i\Omega /2} \ket{10}\bra{10} 
\nonumber \\ 
 & = & \left( \ket{0} \bra{0} + e^{i\Omega/2} \ket{1} \bra{1} \right) 
\nonumber \\ 
 & & \otimes 
\left( \ket{0} \bra{0} + e^{-i\Omega/2} \ket{1} \bra{1} \right)  .  
\label{eq:productgate} 
\end{eqnarray}
Thus, ${\rm U}^{(i)} (\mathscr{C})$ cannot entangle the qubits.

In case (ii), we instead have 
\begin{eqnarray}
{\rm U}^{(ii)} (\mathscr{C}) & \equiv & {\rm U}_{\frac{\pi}{2},\frac{\pi}{2}} (\mathscr{C}) =  
\ket{00} \bra{00} + \ket{11} \bra{11}  
\nonumber \\ 
 & & + e^{-i\Omega /2} \ket{\Psi_+}\bra{\Psi_+}  + 
 e^{i\Omega /2} \ket{\Psi_-}\bra{\Psi_-} 
\end{eqnarray}
with the maximally entangled states $\ket{\Psi_{\pm}} = 
\frac{1}{\sqrt{2}} \big( \ket{01} \pm i\ket{10} \big)$, yielding  
\begin{eqnarray}
{\rm U}^{(ii)} (\mathscr{C}) = \left( \begin{array}{cccc} 
1 & 0 & 0 & 0 \\
0 & \cos \frac{\Omega}{2} & - \sin \frac{\Omega}{2} & 0 \\
0 & \sin \frac{\Omega}{2} & \cos \frac{\Omega}{2} & 0 \\
0 & 0 & 0 & 1 \\
\end{array} \right)  
\label{eq:gengate}
\end{eqnarray}
expressed in the computational basis 
$\{ \ket{00},\ket{01},\ket{10},\ket{11} \}$.

To analyze the entangling capacity of ${\rm U}^{(ii)} (\mathscr{C})$, we calculate 
Makhlin's local invariants $G_1$ and $G_2$ for a two-qubit gate ${\rm U}$ \cite{makhlin02}. 
These are found by the following procedure. First, introduce the unitary operator 
${\rm Q} = \ket{B_1}\bra{00} + \ket{B_2}\bra{01} + \ket{B_3}\bra{10}+\ket{B_4}\bra{11}$ 
that transforms the computational basis into the Bell basis $\ket{B_1} = 
\frac{1}{\sqrt{2}} (\ket{00}+\ket{11})$, $\ket{B_2}=\frac{i}{\sqrt{2}} (\ket{01}+\ket{10})$, 
$\ket{B_3}=\frac{1}{\sqrt{2}} (\ket{01}-\ket{10})$, and 
$\ket{B_4}=\frac{i}{\sqrt{2}} (\ket{00}-\ket{11})$. Second, define 
$m=\left( {\rm Q}^{\dagger} {\rm U} {\rm Q} \right)^{\rm T} {\rm Q}^{\dagger} {\rm U} {\rm Q}$, 
in terms of which the local invariants are found as 
\begin{eqnarray}
G_1 & = & \frac{\Tr^2 m}{16 \det U} ,
\nonumber \\ 
G_2 & = & \frac{\Tr^2 m - \Tr m^2}{4 \det U} , 
\end{eqnarray}
where the latter is real-valued \cite{makhlin02}.

Necessary and sufficient conditions for a perfect entangler (PE) that can maximally 
entangle a product state are $0\leq \left| G_1 \right| \leq \frac{1}{4}$ and 
$-1 \leq G_2 \leq 1$ \cite{balakrishnan10}; a special perfect entangler (SPE) 
\cite{rezakhani04}, which is a gate that maximally entangles an orthonormal product 
basis, is such that $G_1=0$ and $-1 \leq G_2\leq 1$. The SPEs saturate the upper 
bound $\frac{2}{9}$ of the entangling power for qubit pairs \cite{zanardi00}. Direct 
calculation for ${\rm U}={\rm U}^{(ii)} (\mathscr{C})$ yields 
\begin{eqnarray}
G_1 & = & \cos^4 \frac{\Omega}{2}, 
\nonumber \\ 
G_2 & = & 1+2\cos \Omega.
\end{eqnarray}
We thus see that we can only find PEs for solid angles $\frac{\pi}{2} \leq 
\left| \Omega \right| \leq \pi$ with the upper bound corresponding to an SPE. In fact,  
for $\Omega = -\pi$, one finds the {\it i}SWAP-type gate: 
\begin{eqnarray}
{\rm U}^{(ii)} (\mathscr{C}) = \left( \begin{array}{cccc} 
1 & 0 & 0 & 0 \\
0 & 0 & 1 & 0 \\
0 & -1 & 0 & 0 \\
0 & 0 & 0 & 1 \\
\end{array} \right) , 
\label{eq:iswap}
\end{eqnarray}
which is an SPE that maximally entangles the orthonormal product states 
$\frac{1}{2} (\ket{0}\pm\ket{1}) \otimes (\ket{0}\pm\ket{1})$ and $\frac{1}{2} 
(\ket{0}\pm\ket{1}) \otimes (\ket{0}\mp\ket{1})$. 

\begin{figure}[h!]
\centering
\includegraphics[width=0.65\textwidth]{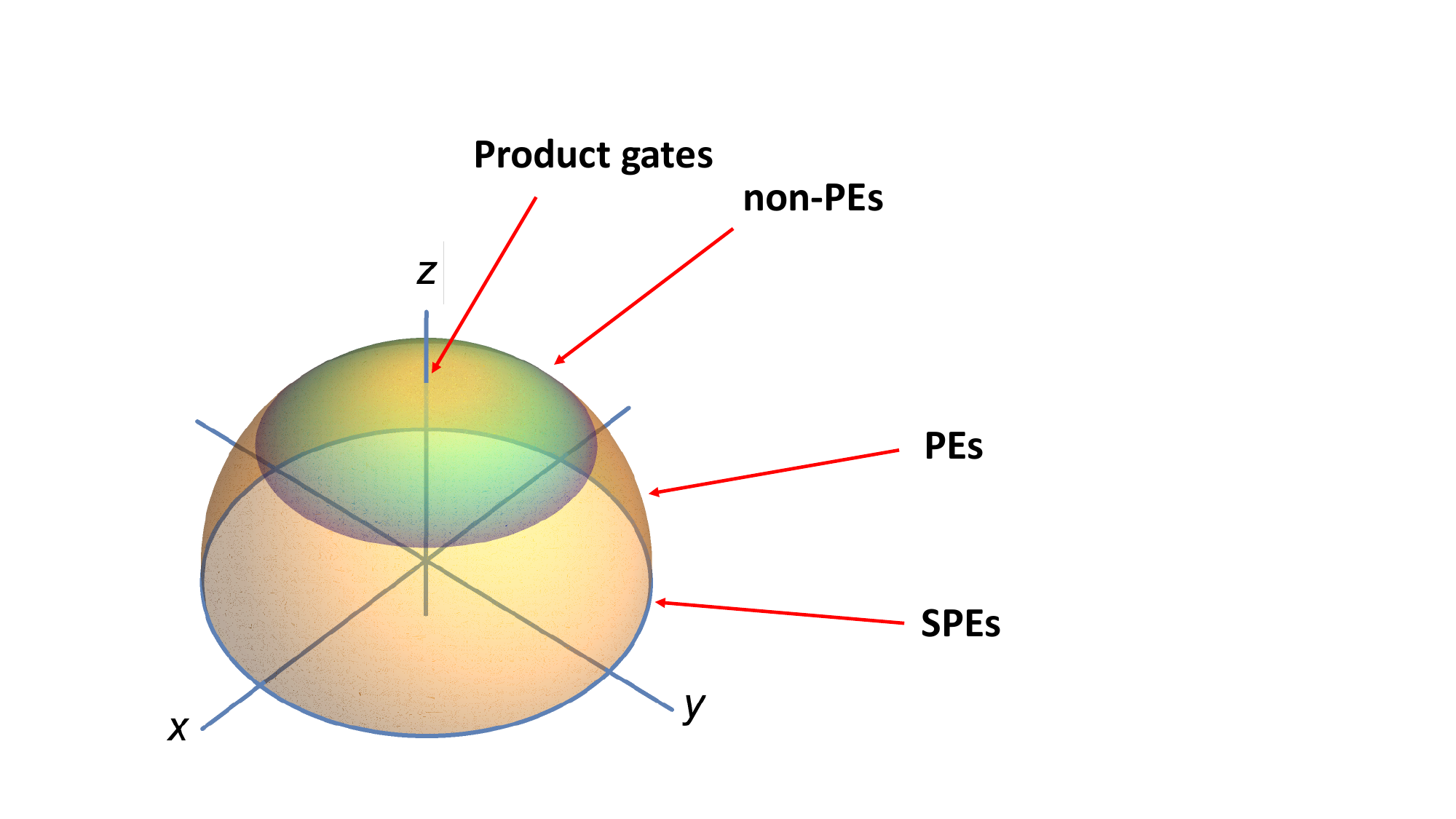}
\caption{Entangling capacity on the Schmidt sphere with $(x,y,z) = 
(\sin \alpha \cos \beta, \sin \alpha \sin \beta, \cos \alpha)$.  Perfect 
entanglers (PEs) are certain loops based at points with polar angles $\alpha_0 
\in [\frac{\pi}{4},\frac{\pi}{2}]$ with 
the special perfect entanglers (SPEs) on the equator $\alpha_0 = \frac{\pi}{2}$. 
There are no product states that can be transformed into a maximally entangled 
state by gates for base points with $\alpha_0 \in [0,\frac{\pi}{2})$. Product gates, 
such as ${\rm U}^{(i)} (\mathscr{C})$ in Eq.~(\ref{eq:productgate}), are located at 
the north pole. While we show only the upper half for clarity, exactly the same 
classification of gates can be found on the lower half of the Schmidt sphere.}
\label{fig:entangling}
\end{figure}

We next analyze which base points ${\bf r}_0$ on the Schmidt sphere allow for 
PEs, as illustrated in Fig.~\ref{fig:entangling}.  For the general 
${\rm U}_{\alpha_0,\beta_0}(\mathscr{C})$, we find the local 
invariants 
\begin{eqnarray}
G_1 & = & \frac{1}{16} \left[ 4 - 2\sin^2 \alpha_0 \left( 1- \cos \Omega \right)\right]^2 , 
\nonumber \\ 
G_2 & = & 3 - 2\sin^2 \alpha_0 \left( 1- \cos \Omega \right) , 
\end{eqnarray}
which confirms that the azimuthal angle $\beta_0$ is irrelevant to the entangling capacity, 
as noted before. We see that there are no PEs for $\alpha_0 \in [0,\frac{\pi}{4})$. We also 
see that $G_1 \geq \cos^4 \alpha_0$, which implies that ${\rm U}_{\alpha_0,\beta_0} 
(\mathscr{C})$ can be an SPE only for base points on the equator $\alpha_0 = \frac{\pi}{2}$.

In the general case, not only $\ket{\Gamma_{\pm}} \in {\rm Span} \{ \ket{01} ,\ket{10}  \}$ 
but also $\Lambda_{\pm} \in {\rm Span} \{ \ket{00} ,\ket{11} \}$ can evolve purely on the 
Schmidt sphere. As the two pairs $\ket{\Gamma_{\pm}}$ and $\ket{\Lambda_{\pm}}$ evolve 
in orthogonal subspaces, such gates factorize into products of commuting geometric 
Schmidt gates: 
\begin{eqnarray}
{\rm U}_{\alpha_0' ,\beta_0' ;\alpha_0 ,\beta_0} (\mathscr{C'},\mathscr{C}) 
& = & {\rm U}_{\alpha_0' ,\beta_0'} (\mathscr{C}') {\rm U}_{\alpha_0 ,\beta_0} (\mathscr{C}) , 
\nonumber \\ 
\left[ {\rm U}_{\alpha_0' ,\beta_0'} (\mathscr{C}'), 
{\rm U}_{\alpha_0 ,\beta_0} (\mathscr{C}) \right] & = & 0. 
\end{eqnarray}
Here, 
\begin{eqnarray}
{\rm U}_{\alpha_0' ,\beta_0'} (\mathscr{C}') & = & \ket{01} \bra{01} + 
\ket{10} \bra{10} 
\nonumber \\ 
 & & + e^{-i \Omega'/2} 
\ket{\Lambda_+ ({\bf r}_0')} \bra{\Lambda_+ ({\bf r}_0')} 
\nonumber \\ 
 & & + e^{i \Omega'/2} 
\ket{\Lambda_- ({\bf r}_0')} \bra{\Lambda_- ({\bf r}_0')} 
\end{eqnarray}
with 
\begin{eqnarray}
\ket{\Lambda_+ ({\bf r}_0')} & = & e^{-i\beta_0'/2} 
\cos \frac{\alpha_0'}{2} \ket{00} + e^{i\beta_0'/2} 
\sin \frac{\alpha_0'}{2} \ket{11} , 
\nonumber \\ 
\ket{\Lambda_- ({\bf r}_0')} & = &  
-e^{-i\beta_0'/2} 
\sin \frac{\alpha_0'}{2} \ket{00} + e^{i\beta_0'/2} 
\cos \frac{\alpha_0'}{2} \ket{11} . 
\end{eqnarray}  
Thus, the only essential difference between ${\rm U}_{\alpha_0' ,\beta_0'} (\mathscr{C}')$ 
and ${\rm U}_{\alpha_0 ,\beta_0} (\mathscr{C})$ is that the former couples $\ket{00}$ and 
$\ket{11}$, while the latter couples $\ket{01}$ and $\ket{10}$); in other words, their 
entangling capacity is the same and thus for both captured by the above analysis.    

\subsection{Reverse engineering}
We now examine the physical realization of the geometric Schmidt gates. We focus on 
the case where only Schmidt vectors in ${\rm Span}\{ \ket{01},\ket{10} \}$ evolve and 
where ${\bf n}_0 = -{\bf m}_0 = (0,0,1)$. Thus, we look for the Hamiltonian that generates 
the time dependent Schmidt vectors  
\begin{eqnarray}
\ket{\Gamma_{+} ({\bf r}_t)} & = & 
f(t)\ket{01} + g(t)\ket{10} , \nonumber \\ 
\ket{\Gamma_{-} ({\bf r}_t)} & = & 
-g^{\ast} (t)\ket{01} + 
f^{\ast} (t) \ket{10} , 
\nonumber \\ 
\ket{\Lambda_{+}} & = &  
\ket{00} , \ \ \ 
\ket{\Lambda_{-}} =  
\ket{11} . 
\label{eq:equator_vectors} 
\end{eqnarray}
To this end, we insert the ansatz 
\begin{eqnarray}
H(t) & = & \omega_{22} (t) \, \ket{01} \bra{01} + \omega_{33} (t) \ \ket{10} \bra{10}   
\nonumber \\ 
 & & + \omega_{23} (t) \ket{01} \bra{10} + {\rm H.c.} 
 \label{eq:gen_ham}
\end{eqnarray}
and Eq.~(\ref{eq:equator_vectors}) into the Schr\"odinger 
equation, yielding ($\hbar = 1$ from now on)
\begin{eqnarray}
i \dot{f} & = & \omega_{22} f + \omega_{23} g ,  \ \ \ 
i \dot{g} = \omega_{23}^{\ast} f + \omega_{33} g , 
\nonumber \\ 
-i \dot{g}^{\ast} & = & -\omega_{22} g^{\ast} + \omega_{23} f^{\ast} ,  \ \ \ 
i \dot{f}^{\ast} = - \omega_{23}^{\ast} g^{\ast} + \omega_{33} f^{\ast} . 
\end{eqnarray}
These equations have the solution 
\begin{eqnarray}
\omega_{22} & = & - \omega_{33} = i\left( \dot{f} f^{\ast} + g \dot{g}^{\ast} \right) ,  
\nonumber \\ 
\omega_{23} & = & i \left( \dot{f} g^{\ast} - f \dot{g}^{\ast} \right) , 
\end{eqnarray}
where we have used that $|f|^2 + |g|^2 = 1$, which in turn implies that 
$ \dot{f} f^{\ast} + g \dot{g}^{\ast} $ is purely imaginary, ensuring that $\omega_{22}$ and 
$\omega_{33}$ are real-valued. By using the parametrization 
$f=e^{-i\beta/2} \cos\frac{\alpha}{2}$ and $g=e^{i\beta/2} \sin \frac{\alpha}{2}$, we find
\begin{eqnarray}
\omega_{22} = -\omega_{33} = \frac{\dot{\beta}}{2}, \ \ \ 
\omega_{23} = -\frac{\dot{\alpha}}{2} (\sin \beta + i \cos \beta) . 
\label{eq:matrix_elements}
\end{eqnarray}
We can use the identities $\ket{0} \bra{0} = \frac{1}{2} (\hat{1} + {\rm Z})$, $\ket{1} \bra{1} = 
\frac{1}{2} (\hat{1} - {\rm Z})$, $\ket{0} \bra{1} = \frac{1}{2} ({\rm X} + i{\rm Y})$, 
and Eq.~(\ref{eq:matrix_elements}) to derive the reverse engineered Hamiltonian 
\begin{eqnarray}
H = -\frac{\dot{\alpha}}{2} \sin \beta \ h_{\rm XY} + \frac{\dot{\alpha}}{2} \cos \beta \ 
h_{\rm DM} + \frac{\dot{\beta}}{2} \ h_{\rm Z} , 
\label{eq:revh}
\end{eqnarray}
where we have identified the XY,  Dzyaloshinskii–Moriya (DM), and Zeeman terms: 
\begin{eqnarray}
h_{\rm XY} & = & \frac{1}{2} \Big( {\rm X} {\rm X} + {\rm Y} {\rm Y} \Big) , \ \ \ 
h_{\rm DM} = \frac{1}{2} \Big( {\rm Y} {\rm X} - {\rm X} {\rm Y} \Big) , 
\nonumber \\ 
h_{\rm Z} & = & \frac{1}{2} \Big({\rm Z} \hat{1} - 
\hat{1} {\rm Z}\Big), 
\end{eqnarray}
respectively. One can verify that these operators satisfy the standard SU(2) algebra: 
$[h_{\rm XY},h_{\rm DM}]=2i h_{\rm Z}$ with cyclic permutations. Thus, $H$ in 
Eq.~(\ref{eq:revh}) describes an effective `spin-$\frac{1}{2}$' ${\bf S} \equiv \frac{1}{2} 
(h_{\rm XY} , h_{\rm DM} , h_{\rm Z})$ interacting with an effective `magnetic field' 
${\bf B} \equiv (-\dot{\alpha} \sin \beta,\dot{\alpha} \cos \beta,\dot{\beta})$.

The paths generated by $H$ are controlled by only two parameters and give rise to 
geometric gates provided the dynamical phases vanish. The latter can be assured 
most easily by following a pair of geodesic segments forming a loop on the Schmidt 
sphere, in analogy with the orange-slice-paths on the Bloch sphere used to implement 
geometric single-qubit gates \cite{thomas11,zhao17,zhou21}. 

\begin{figure}[h!]
\centering
\includegraphics[width=0.7\textwidth]{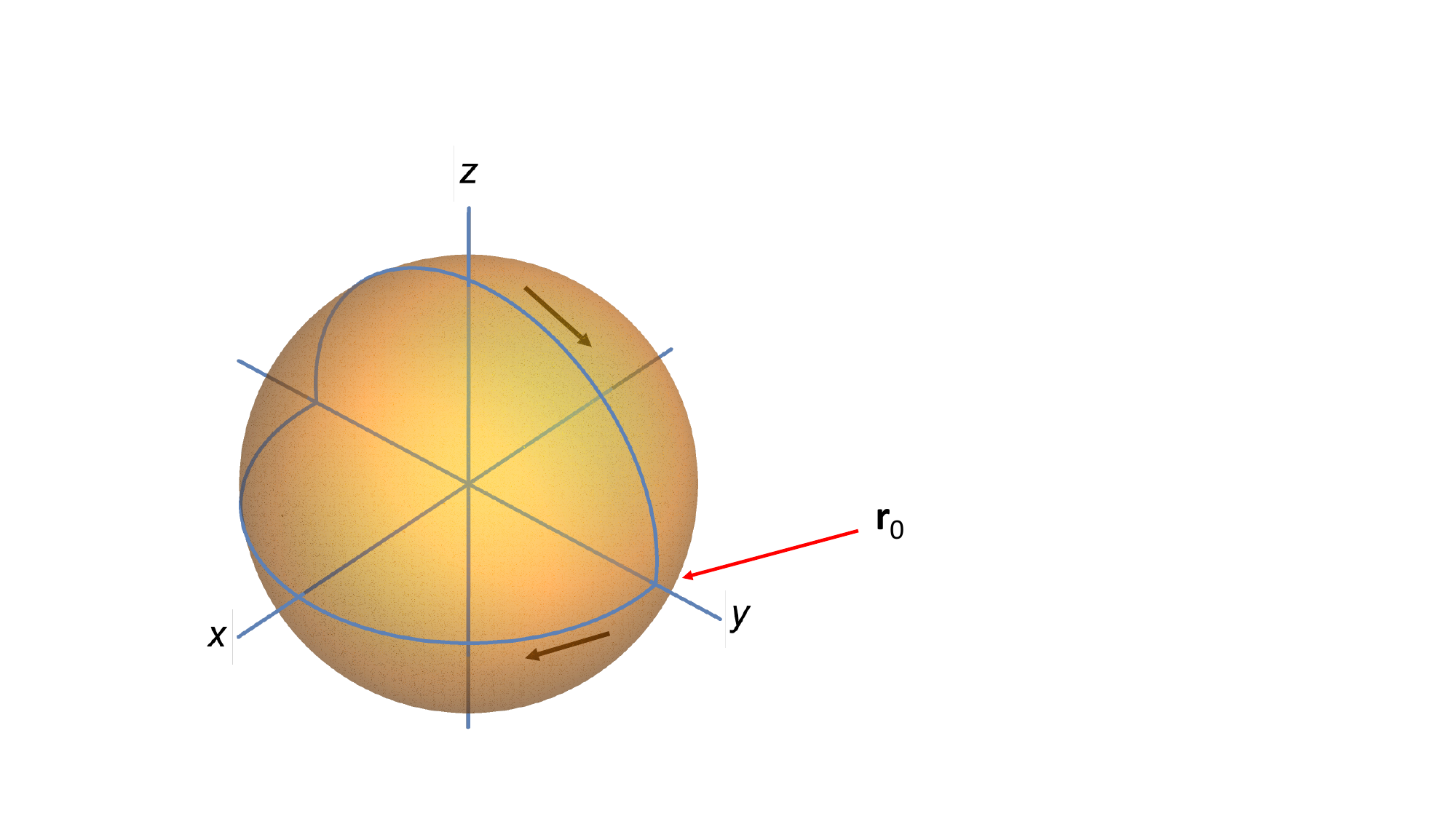}
\caption{Orange-slice curve on the Schmidt sphere that implements the perfect 
entangler ${\rm U}^{(ii)} (\mathscr{C})$ in Eq.~\eqref{eq:iswap}. The curve starts and 
ends at ${\bf r}_0 = (0,1,0)$ and consists of two path segments generated by sequentially 
applying Zeeman and XY interactions to the qubit pair. The enclosed solid angle is $-\pi$.}
\label{fig:polar}
\end{figure}

We illustrate this latter point by demonstrating a realization of the {\it i}SWAP-type gate 
${\rm U}^{(ii)} (\mathscr{C})$ in Eq.~(\ref{eq:iswap}). A geometric implementation of this gate is 
obtained by traversing an orange-slice path on the Schmidt sphere that connects the 
points ${\bf r}_0=(0,1,0)$ and ${\bf r}_{t_1}=(0,-1,0)$ along the equator and thereafter 
back along a geodesic through the north pole: 
\begin{eqnarray}
t & \in & [0,\tau] 
\mapsto {\bf r}_t 
\nonumber \\ 
 & = & \left\{ \begin{array}{cl} 
\left[ \sin \left( \pi \frac{t}{t_1} \right),\cos \left( \pi \frac{t}{t_1} \right),0 \right] , 
& 0 \leq t \leq t_1 , \\
\left[ 0,\cos \left( \pi \frac{t-t_1}{\tau-t_1} \right),\sin \left( \pi \frac{t-t_1}{\tau-t_1} \right) \right], & 
t_1 \leq t \leq \tau, 
\end{array}
\right. 
\end{eqnarray}
thereby enclosing a solid angle $-\pi$, see Fig.~\ref{fig:polar}. This is achieved by 
applying the two-pulse Hamiltonian 
\begin{eqnarray}
H(t) = \left\{ \begin{array}{cl} 
- \frac{\pi}{2t_1} \ h_{\rm Z} , 
& 0 \leq t \leq t_1 , \\
- \frac{\pi}{2\left( \tau - t_1\right)} \ h_{\rm XY}, & t_1 \leq t \leq \tau  . 
\end{array}
\right.
\label{eq:pulseham}
\end{eqnarray}

\subsection{Example: transmon setting}
Our scheme is implementable in any qubit system that naturally include 
XY and DM spin exchange interaction. For instance, the above path that generates   
${\rm U}^{(ii)} (\mathscr{C})$ can be realized for two transmon qubits 
coupled by a transmssion line on a chip \cite{majer07}. Here, the first Zeeman pulse 
in Eq.~(\ref{eq:pulseham}) is generated by detuning the qubits by an amount 
$\pm \delta = \mp \frac{1}{2t_1}$, respectively, from their idle frequencies \cite{salathe15}, 
while the second pulse is realized via exchange of virtual photons in the cavity  by tuning 
the transition frequencies of the two transmon qubits into resonance  \cite{majer07}. 
In this way, one takes advantage of the naturally occuring XY interaction term for 
direct implemention of the geometric {\it i}SWAP-type gate in Eq.~(\ref{eq:iswap}). 

Realization of more general paths in the transmon setting would require simultaneous 
applications of the XY and Zeeman terms. To achieve this, one may use 
Suzuki-Trotter-based techniques \cite{lloyd96} to simulate the effect of such spin 
models \cite{salathe15}. For instance, by performing the second path segment so 
as to make an angle $\theta$ to the $yz$ plane would require application of the 
Hamiltonian 
\begin{eqnarray}
H_{\theta} (t) = 
-\frac{\pi}{2\left( \tau - t_1\right)} 
\left( \cos \theta \ h_{\rm XY} - \sin \theta \ h_{\rm Z} \right) , \ \  t_1 \leq t \leq \tau, 
\end{eqnarray}
which implements ${\rm U}^{(ii)} (\mathscr{C})$ in Eq.~(\ref{eq:gengate}) with $\Omega = 2\theta$. 
This can be simulated by performing 
\begin{eqnarray}
U_n = e^{i\frac{\pi}{2n} \cos \theta \ h_{\rm XY}} e^{-i\frac{\pi}{2n} \sin \theta \ h_{\rm Z}}
\end{eqnarray}
$n$ times. In this way, the solid angle dependence of the gate can be tested by 
varying $\theta$. 

\section{Conclusions}
In conclusion, we have demonstrated a class of two-qubit gates associated with paths 
on the Schmidt sphere. These gates control the entangling capacity of the two-qubit 
evolution and have a clear geometric interpretation in terms of solid angles enclosed on 
the Schmidt sphere. A key point of our proposal is that it provides means for experimentally 
implementing {\it i}SWAP-type geometric gates based on spin exchange terms that naturally 
appear in several qubit architectures. 

To complete a universal set, one needs to implement sufficiently flexible single-qubit 
gates. One may achieve this by means of paths on the local Bloch spheres, while 
keeping the Schmidt parameters $\alpha,\beta$ fixed. These gates can be assured 
to be geometric by designing the paths so that the dynamical phases all vanish, for 
instance by using geodesic segments on the Bloch spheres. Thus, by generating 
ordered sequences of paths on Schmidt and Bloch spheres, any quantum computation 
can be realized efficiently by purely geometric means.


\begin{thebibliography}{99}
\bibitem{remark1} More precisely, for $|f|=|g|$, only one of the Bloch vectors ${\bf n}$ 
and ${\bf m}$ determines uniquely the state as all maximally entangled states can be 
reached by locally manipulating only one of the qubits \cite{milman06}.
\end{thebibliography}
\end{document}